\documentclass[%
reprint,
superscriptaddress,
pra,
]{revtex4-2}

\usepackage{graphicx} 
\usepackage{amsmath} 
\usepackage{xcolor}
\usepackage{braket}
\usepackage[colorlinks=true, linkcolor=black, anchorcolor=black, citecolor=blue,
urlcolor = blue]{hyperref}
\usepackage[autostyle]{csquotes}
\newcommand{\mycomment}[1]{}
\usepackage{soul}
\usepackage{lipsum}
\usepackage{braket}
\usepackage{siunitx}
\usepackage{xr}
\usepackage{soul}
\begin{document}

\title{Achieving a long lifetime synchronized state in a thermal vapor Rydberg time crystal}

\author{William J. Watterson}
\email{Corresponding author: william.watterson@nist.gov}
\affiliation{National Institute of Standards and Technology, Boulder, CO 80305, USA}
\author{Dixith Manchaiah}
\affiliation{National Institute of Standards and Technology, Boulder, CO 80305, USA}
\affiliation{Department of Physics, University of Colorado, Boulder, Colorado 80309, USA}
\author{Christopher L. Holloway}
\affiliation{National Institute of Standards and Technology, Boulder, CO 80305, USA}

\date{\today}

\begin{abstract}

Time crystals are a unique and recently discovered state of matter in which time-translational symmetry is spontaneously broken. Although this state has been observed across a variety of quantum systems, understanding the nature between driving, dissipation, and noise and the effects of these forces on the state lifetime remains elusive. Here, we study phase diffusion and entrainment of the time crystalline oscillatory state of a thermal-vapor Rydberg atom ensemble. We apply a pulsed rf heterodyne driving field which can entrain the oscillation frequency during the on driving period. During the off period, the entrained ensemble phase diffuses on a 1/$e$ timescale up to 2.3 ms -- significantly longer than the Rydberg state lifetime. We additionally study how the entrainment and phase diffusion timescales are affected by the strength and detuning of the rf driving field, and develop a classical Fokker-Planck model with an Adler-based driving term that reproduces the observed experimental findings. These results demonstrate the creation of a long-time synchronized time crystal Rydberg atom ensemble state, which could be utilized in applications requiring frequency and phase stability. 

\end{abstract}

\maketitle

\section{Introduction}
Symmetry breakings and their associated phase transitions are a ubiquitous phenomena found throughout nature, with everyday examples including spatial symmetry breaking underlying the liquid to solid phase transition and spin rotational symmetry breaking forming the ferromagnetic phase. In 2012, Wilczek and Shapere formulated an intriguing analogous conjecture that time translational symmetry breaking could lead to a so-called time crystal phase in which late-time quantum observables exhibit persistent oscillations, ergodicity breaking, and a lack of thermalization \cite{shapere2012classical, wilczek2012quantum}. Although their initial formalization related to equilibrium ground state systems was shown to be theoretically incompatible with a time crystal phase \cite{watanabe2015absence}, a host of experimental results demonstrated that emergent synchronization in non-equilibrium driven-dissipative interacting systems can indeed support persistent oscillations in quantum observables \cite{zaletel2023colloquium, khemani2019brief, else2020discrete, zhang2017observation}. These states exist for both discrete and continuous time symmetry breaking -- where the discrete time crystal (DTC) features a sub-harmonic response to external periodic forcing \cite{zhang2017observation, else2016floquet, frey2022realization, else2020discrete, yao2020classical, kyprianidis2021observation} while the continuous time crystal (CTC) develops persistent oscillations under constant driving \cite{kongkhambut2022observation, iemini2018boundary, kozin2019quantum, greilich2024robust, pohl2025quantum}.

One open question and active field of research concerns the stability of the oscillation's frequency and phase in the time-crystalline order \cite{zaletel2023colloquium, zhang2017observation, kongkhambut2022observation, jiao2025photoionization}. If noise in the system induces phase slips at a rate 1/$\tau$, then the system will have an autocorrelation time of $\tau$ \cite{zaletel2023colloquium}. This phase noise manifests as a distribution in the natural oscillation frequency and arises due to finite strength coupling between the interacting constituents when at non-zero temperature. Gaining a further understanding of the stability of the time crystal phase, and what parameters can be tuned to manipulate it, could have important implications for proposed time crystal applications requiring frequency and phase stability, including metrology \cite{yousefjani2025discrete, liu2026enhanced, iemini2024floquet, arumugam2025electric}, quantum computing \cite{frey2022realization, mi2022time}, and timing \cite{viotti2026quantum}.

Here, we utilize a hot vapor Rydberg atom time crystal phase to study the ensemble oscillatory phase dynamics under pulsed radiofrequency (rf) driving. The Rydberg atom time crystal is formed by exciting alkali atoms to a high principal quantum number, $n$, in a three-level ladder configuration with strong driving laser fields to induce many-body dynamics mediated by van der Waal's forces (with $C_6$ coefficients scaling as $n^{11}$). This interplay between constant laser driving and dissipation can lead to a stable limit cycle and synchronized oscillations under certain driving parameters \cite{wadenpfuhl2023emergence, wu2024dissipative, ding2024ergodicity, jiao2025observation} -- consistent with the interpretation of continuous time crystal state.

The dynamics can be further enriched by applying a radiofrequency field to couple nearby Rydberg states \cite{liu2025bifurcation, liu2024higher, arumugam2026injection, manchaiah2026frequency, wu2026dissipative} -- interesting observations have shown complex non-equilibrium phenomena including bifurcation of the time crystal phase \cite{liu2025bifurcation}, fractional time crystal orders \cite{liu2024higher}, injection locking \cite{arumugam2026injection}, and frequency comb formation \cite{manchaiah2026frequency}. In certain cases, for example under injection locking of the CTC Rydberg state, the oscillation frequency interacts with the rf field in a classical manner of coupled oscillators \cite{arumugam2026injection}. However, in other instances, the CTC Rydberg state under Floquet driving can develop a subharmonic periodic response of the driving frequency, and therefore features behaviors more consistent with DTC states \cite{liu2024higher}. 

Regardless of the drive parameters, noise present in the bath inevitably couples to dissipative interactions of the oscillations -- with many studies frequently observing phase diffusion and a Lorentzian distribution of the natural oscillation frequency. By introducing an rf signal, the rf forcing may overcome this noise under strong driving parameters. Here, we study this interaction between driving, dissipation, and noise using pulsed RF driving to study the time crystal entrainment and decay time dynamics.

\section{Methods}

\begin{figure}[htbp!]
    \includegraphics[width=0.95\columnwidth]{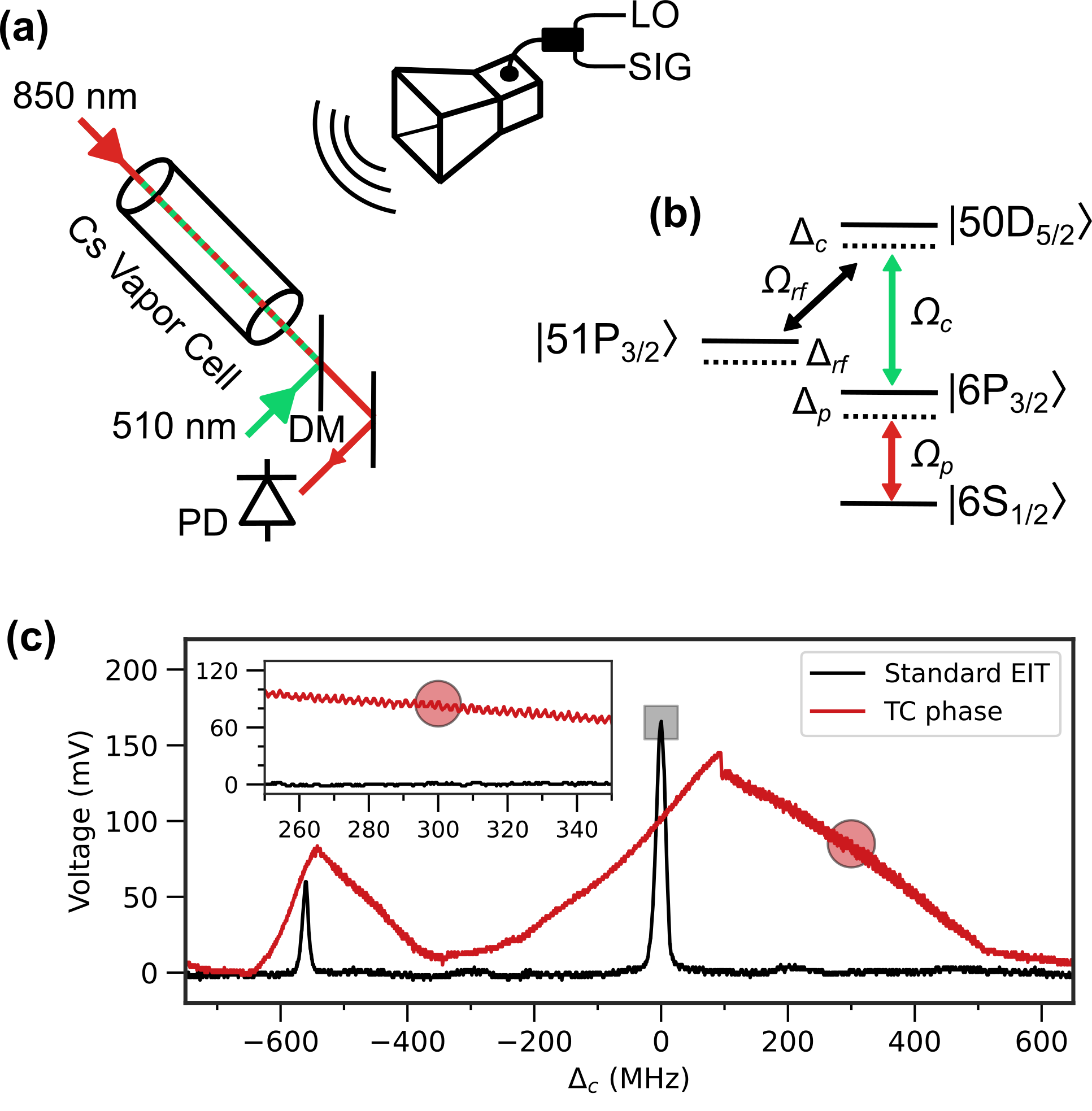}
    \caption{\textbf{Experimental setup.} (a) Counter-propagating laser fields excite a vapor of Cs atoms to a Rydberg state. The LO and SIG fields are combined via a power combiner and transmitted from the same horn. DM: Dichroic mirror. PD: Photodiode. (b) Cs energy level diagram. (c) Spectra of a standard EIT and the time crystal phase scanned over the coupling detuning, $\Delta_c$. The black and red dots indicate the coupling frequency lock location for time-domain signals plotted in Fig. \ref{Fig:TCdynamics}, respectively. No RF is applied in these spectra. The inset shows a zoomed in display of the region highlighted by the red circle.}
    \label{Fig:Experiment}
\end{figure}

We first excite cesium atoms to the Rydberg state using a two-photon excitation scheme (Fig. \ref{Fig:Experiment}a) to create an electromagnetically induced transparency (EIT) in the atomic spectra. See, for example, Ref. \cite{holloway2014broadband} for an in-depth discussion on EIT. In particular, we use an 850 nm diode laser for the probe beam and a counter-propagating 510 nm coupling beam derived from a 1020 nm fundamental via second-harmonic generation. The probe response is recorded on a photodiode with 8 MHz bandwidth. We show the probe response while scanning the coupling laser detuning, $\Delta_c$, for both the case of a standard EIT (sEIT) and for the time crystal (TC) phase (Fig. \ref{Fig:Experiment}c). In the case of the standard EIT, we drive the transition $\ket{6\mathrm{S}_{1/2}, F=4} \rightarrow \ket{6\mathrm{P}_{3/2}, F' = 5} \rightarrow \ket{50\mathrm{D}_{5/2}}$. For the time crystal phase, we only experimentally observe oscillations when the probe is detuned from the $F'=4$ or $F'=5$ state. Here, we choose to excite from  $\ket{6\mathrm{S}_{1/2}, F=4} \rightarrow \ket{6\mathrm{P}_{3/2}, F' = 4,5} \rightarrow \ket{50\mathrm{D}_{5/2}}$. The laser Rabi rates in Fig.~\ref{Fig:Experiment}b are $\Omega_p^{sEIT}/2\pi =$ 62~MHz, $\Omega_c^{sEIT}/2\pi =$ 2.8~MHz, $\Omega_p^{TC}/2\pi =$ 130~MHz, and $\Omega_c^{TC}/2\pi =$ 28~MHz. The vapor cell is 25 cm diameter by 75 cm in length and is insulated and heated to 50\si{\degreeCelsius} for the time crystal phase using two ceramic heaters located at the vapor cell windows. The cell is kept at room temperature in the case of the standard EIT. Fig. \ref{Fig:Experiment}c shows the standard EIT of $\ket{50\mathrm{D}_{5/2}}$ along with the fine structure component of $\ket{50\mathrm{D}_{3/2}}$, which is used for frequency calibration. In the high Rabi regimes, we observe spectral signatures characteristic of emergent synchronization including an optical bistability transition and a region showing persistent oscillations (shown in red). \cite{wadenpfuhl2023emergence, manchaiah2026frequency}.


RF fields are introduced under a heterodyne configuration \cite{simons2019rydberg} to address the transition $\ket{50\mathrm{D}_{5/2}} \rightarrow \ket{51\mathrm{P}_{3/2}}$. Heterodyned Rydberg reception mixes a local oscillator RF field (LO) and a signal rf field (SIG) to down-convert from the carrier frequency to the low-frequency beatnote given by $\Delta \omega_{bn} = \omega_{LO} - \omega_{SIG}$. The carrier frequencies are set to $f_{RF}^{sEIT} = 5.720$ GHz for the standard EIT and $f_{RF}^{TC} = 6.100$ GHz for the TC phase. The LO power is chosen to maximize the decay time and is set to $P_{LO}^{EIT}=$ 0 dBm and $P_{LO}^{TC}=$ 5 dBm. The field propagates orthogonally to the laser direction and is co-polarized with the laser fields. To lock the phases between the two RF signals, two signal generators are clocked to a 10 MHz reference and the pulse period is set to an integer multiple of the beatnote frequency.

\begin{figure*}[htbp!]
    \includegraphics[width=\linewidth]{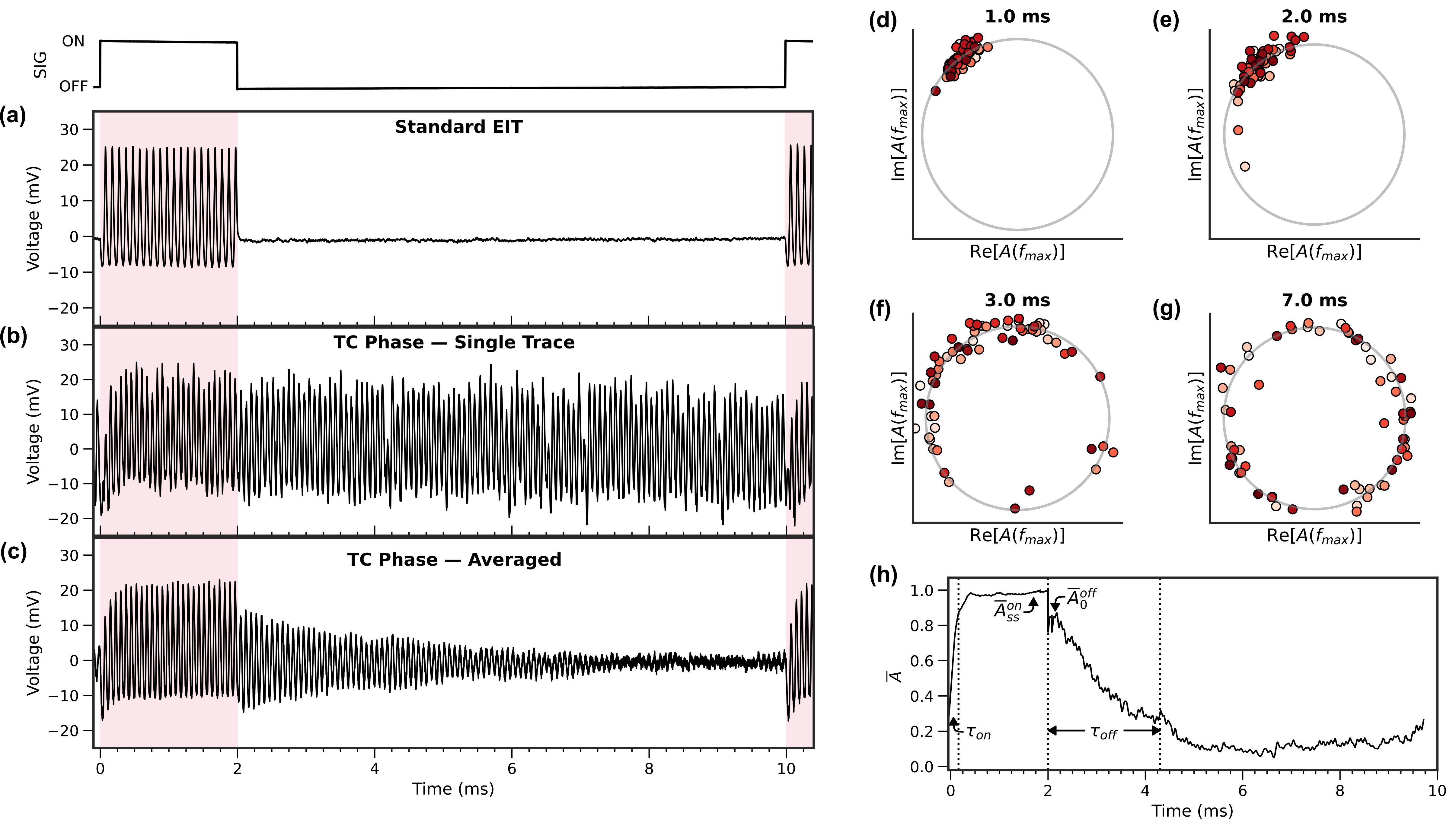}
    \caption{\textbf{Phase entrainment and dephasing in a hot vapor Rydberg time crystal under pulsed RF modulation.} Pulsed SIG field with 2 ms on and 8 ms off time applied to (a) a standard EIT configuration and (b and c) the TC phase. The LO is left continuously on in all cases. (a) A heterodyned beatnote occurs when both SIG and LO are on, followed by a $1/e$ decay time of 14 $\mu$s to a non-modulated state. (b) A single oscilloscope trace of the TC state (c) 64 trace averages over the TC state. (d-g) Fourier components obtained from each of 64 single traces of the TC phase at the time indicated above each plot, with each dot representing one captured trace and the color representing the trace index. (h) The normalized envelope for the averaged scope trace with the following variables indicated: rise time, $\tau_{on}$; decay time $\tau_{off}$; steady state normalized amplitude during the on phase, $\bar{A}^{on}_{ss}$; and normalized initial off time amplitude, $\bar{A}^{off}_{0}$.}
    \label{Fig:TCdynamics}
\end{figure*}

In order to analyze the phase dynamics of the time crystalline order under pulsed rf heterodyned driving, we capture 64 traces from an oscilloscope at each independent experimental condition. We then average the 64 observations, extract the oscillation envelope using a Hilbert transform, and fit the rise time, decay time, and amplitudes in response to the pulsed driving field. Each experiment is repeated a minimum of 4 times in order to determine variation in response. Error bars are presented as the standard deviation over the independent experimental repeats. An example is shown in Fig. \ref{Fig:TCdynamics} and discussed in depth in the Results section.

\section{Model}

The Adler equation was recently proposed as a classical model capable of describing injection locking of a Rydberg time crystal phase under rf driving \cite{arumugam2026injection}. In this model, the author considered an effective reduced closed V-scheme 3-level atomic model, where the intermediate state is eliminated, and where the ground state, $\ket{g}$, is coupled to two Rydberg states, $\ket{r}$ and $\ket{s}$. This effective model was shown to reproduce the observed experimental results, although in reality, the author used a similar two-photon transition to reach the Rydberg state as we do. Here, we extend this model to include noise terms and show that this classical picture captures most of the observed phase dynamics. 

We begin by briefly describing the model given in Ref. \cite{arumugam2026injection} before considering the effect of adding noise. In this model, if the coherence $\sigma_{rs}$ is expressed as 

\begin{equation}
    \sigma_{rs} = A(t)e^{i(\omega_{bn}t+\phi(t))},
\end{equation}

\noindent with $A(t)$ the oscillation amplitude, $\omega_{bn}$ the (beatnote) drive frequency, and $\phi(t)$ the phase with respect to the driving field, then the phase equation can be mapped from the optical Bloch equations as 

\begin{equation}
\begin{split}
    \dot{\phi} &= (\omega_{osc} - \omega_{bn}) -\frac{\Omega_{rs}}{2A}(n_r - n_s) \sin{\phi} \\
     &\equiv\Delta\omega - K\sin\phi,
    \label{Eq:Adler}
\end{split}
\end{equation}

\noindent where $\omega_{osc}$ is the natural oscillation frequency of the time crystal phase, $\Omega_{rs}$ is the rf Rabi frequency, and $n_r-n_s$ is the Rydberg state population inversion. The frequency detuning, $\Delta\omega \equiv \omega_{osc} - \omega_{bn}$, and the forcing strength $K \equiv \frac{\Omega_{rs}}{2A}(n_r - n_s)$. Equation (\ref{Eq:Adler}) has the same form as the Adler equation in classical dynamics. This model gives the injection locking frequency $\Delta\omega \leq K$.

We show that the collective excitation and decay dynamics to pulsed rf driving can be well modeled through the addition of a noise term. The Adler equation with a Langevin noise, $\xi(t)$, can be expressed as

\begin{equation}
    \dot{\phi} = (\omega_{osc} -  \omega_{bn})  - \frac{\Omega_{rs}}{2A}(n_r - n_s) \sin{\phi} + \xi{(t)},
    \label{Eq:AdlerWithNoise}
\end{equation}

\noindent where $\langle \xi(t)\xi(t+\tau)\rangle = \sigma^2 \delta(t)$. Here, $\delta(t)$ is the Dirac delta and $\sigma^2$ is the noise intensity. We note that the noise term, $\xi(t)$, is frequently employed across classical systems and adding it in here gives results which are phenomenologically consistent with our experimental data. However, it is important to note that this does not establish a systematic link between the relevant quantum parameters and the noise variance, $\sigma^2$. When the rf driving is turned off, the unwrapped phase equation (\ref{Eq:AdlerWithNoise}) is then just a random walk with drift given by

\begin{equation}
    \dot{\phi} = \omega_{osc} + \xi{(t)}.
    \label{Eq:Phidot}
\end{equation}

\noindent Without an applied rf, the phase is just defined against an initial time, rather than against the rf driving frequency. Eq. (\ref{Eq:Phidot}) can by solved as

\begin{equation}
    \phi(t) = \phi_0 + \omega_{osc}t + \int_0^t{\xi{(t')}dt'}.
    \label{Eq:PhiTimeSolution}
\end{equation}

\noindent For delta-correlated Gaussian noise, the phase therefore follows a normal distribution given by $\phi(t) \sim \mathcal{N}(\phi_0+\omega_{osc} t, \sigma^2t)$. From Eq. \ref{Eq:PhiTimeSolution}, and using the Wiener–Khinchin theorem, the power spectra, $S(\omega)$, of the oscillations without rf driving can be shown to be \cite{pikovsky2003universal}

\begin{equation}
\begin{split}
    S(\omega) &= \int_{-\infty}^{\infty}\langle \cos(\phi(t))\cos(\phi(t+\tau))\rangle e^{i\omega\tau}d\tau\\
    &= \frac{1}{2}\frac{\frac{1}{2}\sigma^2}{(\frac{1}{2}\sigma^2)^2 + (\omega - \omega_{osc})^2}.
\end{split}
\end{equation}

\noindent Indicating that noise in the non-driven oscillatory state has a Lorentzian distribution with a full width at half max, FWHM = $\sigma^2/2\pi$, in units of cycles/s. Similarly, the expected waveform can be shown to be

\begin{equation}
    \langle\sin(\phi(t))\rangle = e^{-\sigma^2t/2}\sin(\omega_{osc}t+\phi_0).
\end{equation}

\noindent Therefore, the average decay time, $\tau_{off}$ is directly related to the FWHM by $\tau_{off} = 2/\sigma^2$ = 1/($\pi$FWHM). 

Next, to analyze the time dynamics under rf driving, we consider a Fokker-Planck formulation with a deterministic Adler-based drift term according to 

\begin{equation}
    \frac{\partial P(\phi,t)}{\partial t} = -\frac{\partial}{\partial \phi}[(\Delta \omega - K\sin \phi)P(\phi,t)] + \frac{\sigma^2}{2}\frac{\partial^2 P(\phi,t)}{\partial \phi^2},
    \label{Eq:FokkerPlanck}
\end{equation}

\noindent where $P(\phi,t)$ is the probability to find the phase at some value, $\phi$, at a given time, $t$. At zero detuning, the steady state, $\partial P_{ss}(\phi,t;\Delta\omega=0) / \partial t = 0$, is given by

\begin{equation}
    P_{ss}(\phi) = \frac{\exp(\frac{2K}{\sigma^2}\cos\phi)}{2\pi I_0(\frac{2K}{\sigma^2})},
\end{equation}

\noindent where $I_j$ is the modified Bessel function of the first kind of order $j$. The steady state expected waveform is 

\begin{equation}
\begin{split}
    \langle\sin(\omega_{osc}t + \phi)\rangle &= \int_{-\pi}^{\pi}\sin(\omega_{osc}t + \phi) P_{ss}(\phi)d\phi\\
    &= \frac{I_1(\frac{2K}{\sigma^2})}{I_0(\frac{2K}{\sigma^2})}\sin(\omega_{osc}t).
\end{split}
\end{equation}

\noindent The factor $I_1(2K/\sigma^2)/I_0(2K/\sigma^2)$ gives the normalized averaged amplitude and shows a simple dependence on the driving strength $K$ versus noise intensity, $\sigma^2$. To fit the data, we assumed the population inversion, $n_r-n_s \propto \Omega_{rs}$, so that $K  \propto \Omega_{rs}^2 \propto P$. Given the agreement between our model and experimental data, we are therefore likely in a regime where the Rabi rates, detunings, and decay rates give a population inversion proportional to the RF Rabi. Next, since $\sigma^2$ is easily measured from the FWHM off a spectrum analyzer and is independent of $P$, we therefore only use one fit parameter; where we use $K$ as the fit parameter.

Next, we consider how the expected oscillation rise time varies when turning an rf field from off to on depends on $\Omega_{rs}$. We begin by applying separation of variables with $P(\phi,t) = T(t)\Phi(\phi)$ to the Fokker-Plank equation (\ref{Eq:FokkerPlanck}), set $\Delta\omega=0$, and make the change of variables $\phi \rightarrow 2\phi$, to get 

\begin{equation}
    \frac{\partial T}{\partial t} = -\lambda_n T
    \label{Eq:FokkerPlanckT}
\end{equation}
\begin{equation}
    \frac{\sigma^2}{8}\frac{\partial^2 \Phi}{\partial \phi^2} + \frac{K}{2}\sin2\phi\frac{\partial \Phi}{\partial \phi} + (K \cos 2\phi + \lambda_n)\Phi = 0,
    \label{Eq:FokkerPlanckPhi}
\end{equation}

\noindent where $\lambda_n$, $n=1,2,3,..$, are real eigenvalues which give periodic solutions to $\Phi$. From the first equation, we see the leading order rise time ($n$=1) is given by $\tau_{on} = 1/\lambda_1$. The equation for $\Phi$ is known as Ince's equation \cite{arscott2014periodic}. The eigenvalues can be determined by substituting $\Phi(\phi) = \sum_{n=0}^\infty a_n \cos(n\phi)$ or $\Phi(\phi) = \sum_{n=1}^\infty b_n \sin(n\phi)$ and solving a tri-diagonal recurrence relation. The asymptotic eigenvalues under weak and strong driving are given by, respectively, as \cite{arscott2014periodic}

\begin{equation}
    \lambda_n|_{K\ll\sigma^2/2} = \sigma^2n^2 / 2
\end{equation}

\noindent and

\begin{equation}
    \lambda_n|_{K\gg\sigma^2/2} = Kn.
\end{equation}

\noindent The leading order contribution for the rise time ($n$=1) under weak driving is therefore $\tau_{on} =  2/\sigma^2 = \tau_{off}$. And, at strong driving, $\tau_{on} \propto 1/\Omega_{rs}^2$. To fit the experimental data to the rise time $\tau_{on}$, we plot $\beta/\lambda_n(\alpha)$, where $\alpha$ and $\beta$ are free fit parameters. In a similar manner to the steady state amplitude, the fit parameter $\alpha$ is required because we do not have direct access to the proportionality factor in $K$. We include the free fit parameter $\beta$ instead of using $\tau_{off}$ because the solution to Eqs. (\ref{Eq:FokkerPlanckT}) and (\ref{Eq:FokkerPlanckPhi}) consists of a superposition of terms and the leading order decay was not sufficient to fit the data.

\begin{figure*}[htbp!]
    \includegraphics[width=0.9\linewidth]{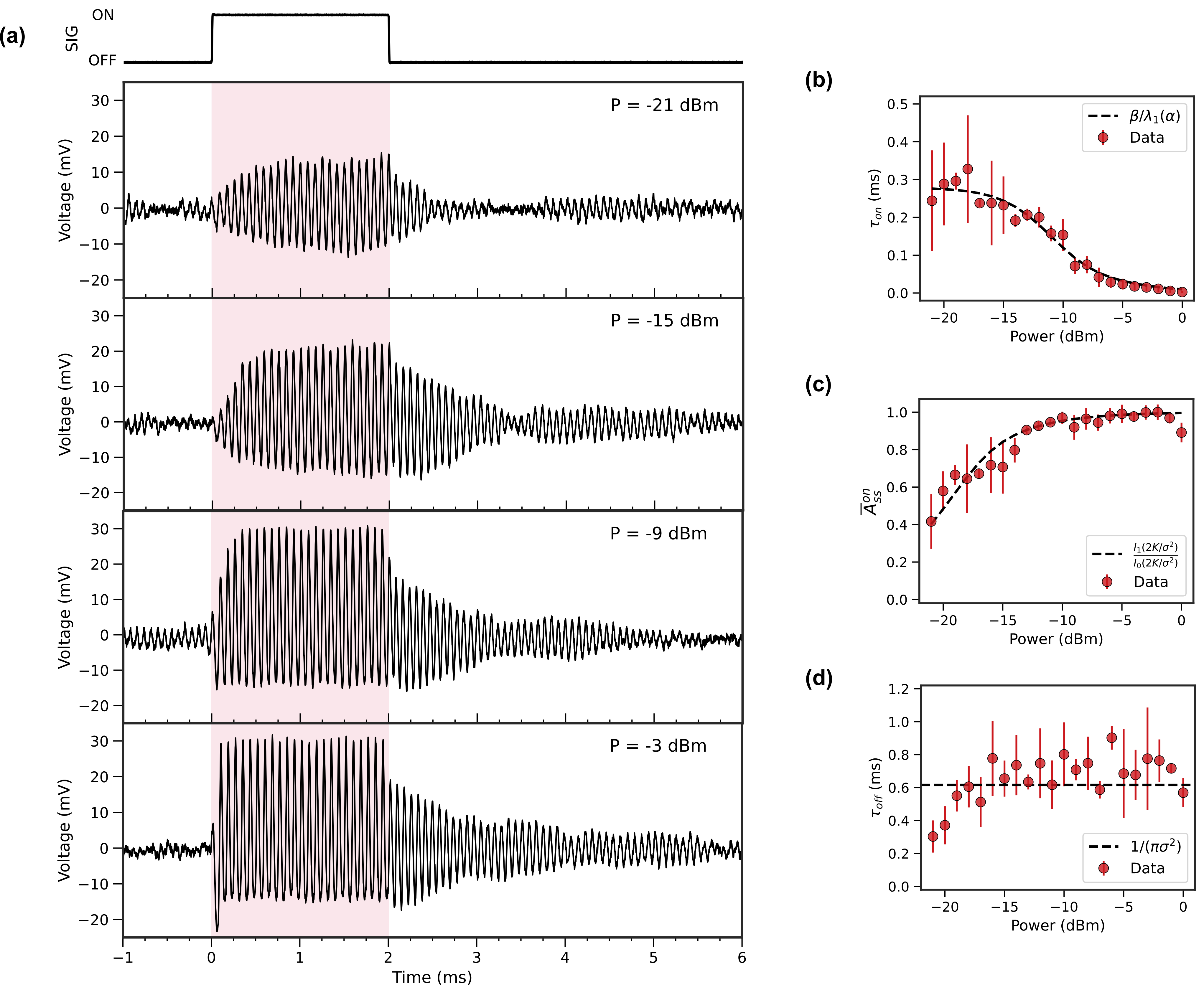}
    \caption{\textbf{Effect of SIG power at constant LO and detuning.} (a) Averaged (64x) time domain TC phase oscillations under SIG pulsing at increasing SIG power and fixed beatnote and LO power. (b) Fitted rise time. (c) Normalized amplitude at the end of the on pulse. (d) Averaged decay time. For the experiments in this figure, the measured FWHM was $\sigma^2 = 516$ Hz.}
    \label{Fig:PowerSweep}
\end{figure*}

\begin{figure*}[htbp!]
    \includegraphics[width=0.85\linewidth]{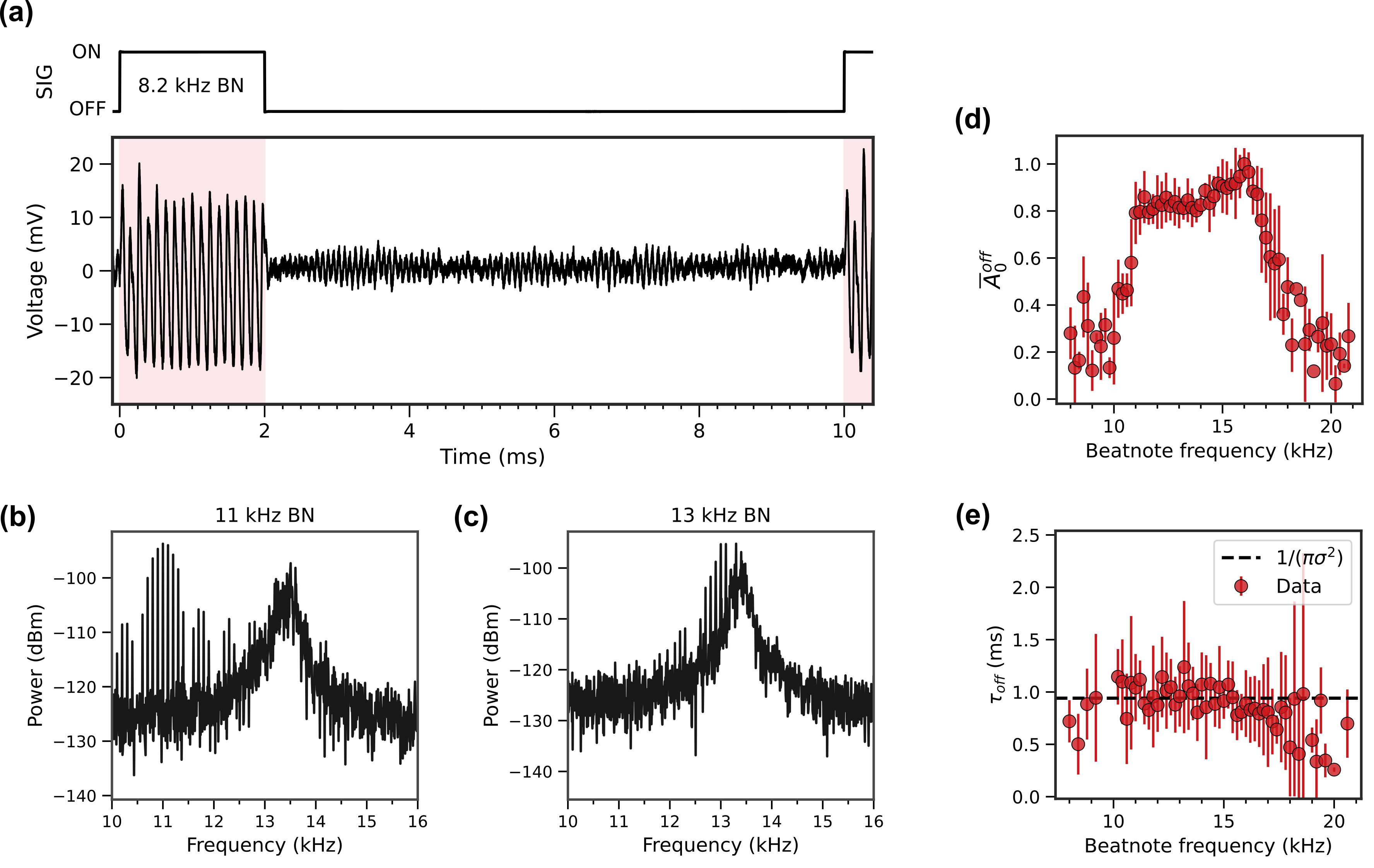}
    \caption{\textbf{Effect of beatnote detuning at constant LO and SIG power.} (a) Averaged (64x) time domain TC phase oscillations at a 13.5 kHz natural oscillation frequency and driven with an 8.2 kHz beatnote during the SIG on pulse. Power spectra of the atomic oscillation and pulsed SIG field for a (b) 11 kHz and (c) 13 kHz beatnote. (d) Order parameter at the start of the SIG off time plotted against beatnote detuning. (e) Decay time against beatnote detuning. The measured FWHM was $\sigma^2 = 318$ Hz.}
    \label{Fig:Detuning}
\end{figure*}


\section{Results}

To compare the time dynamics when introducing the two rf fields,  a constant LO and a pulsed SIG field, to the standard EIT and time crystal phases, we first lock the coupling laser to the locations indicated in Fig. \ref{Fig:Experiment}c. In the case of SIG pulsing on a standard EIT configuration, we observe a 1/$e$ decay time of 14 $\mu$s (Fig. \ref{Fig:TCdynamics}a). In general, the total EIT decay rate, $\gamma_T$, is set by the decay time of the Rydberg state and includes collisional, radiative decay, and black-body radiative state transfer effects \cite{prajapati2024investigation}.   

The intrinsic Rydberg time crystal oscillation frequency for three-level excitation is typically reported to be tunable within a few kHz up to a few tens of kHz. Here, we consider an example with the intrinsic oscillation frequency near 13 kHz for the example shown Fig. \ref{Fig:TCdynamics}b-c. We apply a constant LO field and pulse a SIG field detuned 13.2 kHz from the LO, so that the heterodyned beatnote frequency is near the intrinsic oscillation frequency. We note that previous research has demonstrated that continuous (non-pulsed) rf driving near the intrinsic oscillation frequency can enable injection locking such that the natural oscillation frequency becomes locked to the driving frequency \cite{arumugam2026injection, manchaiah2026frequency}. Over a single trace, the atomic coherence can be seen to remain in a oscillatory state with roughly constant amplitude during and after application of the SIG field (Fig. \ref{Fig:TCdynamics}b). As discussed in the introduction, the TC oscillations feature phase instability and drift, which is uniformly distributed over the interval [0, 2$\pi$] for sufficiently long times. Therefore, by averaging over multiple traces, the averaged TC amplitude will go to zero if the phases are fully diffuse because the average of sine waves with random phase will go to zero as the number of averages goes to infinity. Conversely, if the system retains some amount of phase entrainment, the averaged amplitude will not go to zero. Fig. \ref{Fig:TCdynamics}c shows an average across 64 independent acquisitions and demonstrates entrainment of the oscillatory phase with a timescale $\tau_{on} = $ 0.16 ms when turning the SIG field on, followed by a slow decay of $\tau_{off} = $ 2.3 ms over which the time crystal oscillations dephase. During the SIG on time, the atomic ensemble features a population of atoms interacting in a time crystal state and phase locked by the rf drive as well as a population in a standard heterodyned EIT state, as indicated by the initial drop in the averaged voltage at $t=2$ ms in Fig. \ref{Fig:TCdynamics}c. In Fig. \ref{Fig:TCdynamics}d-g, the phase drift of the fundamental is observed via the Fourier components acquired for each of the 64 traces at four different times. The phase acquired across different scope acquisitions does not appear to drift linearly with acquisition time as indicated by the random distribution of phase verses acquisition number. The extracted parameters from the oscillation envelope are shown in \ref{Fig:TCdynamics}h.

We next explored the effect of varying the SIG power on the TC synchronization and phase dynamics at constant LO power and detuning (Fig. \ref{Fig:PowerSweep}). At the onset of the SIG field, the experimentally measured rise time, $\tau_{on}$, can be seen to agree with our classical Fokker-Planck model with Adler-based drift (Fig. \ref{Fig:PowerSweep}b). Although presumably, the strong driving limit, $\tau_{on} \propto 1/P$, cannot decrease past the standard EIT decay time, but this is not captured in the classical model. At the end of the 2 ms on pulse, the steady state normalized amplitude $\bar{A}_{ss}^{on} = I_1(2K/\sigma^2)/I_0(2K/\sigma^2)$ (Fig. \ref{Fig:PowerSweep}c). This suggests at low SIG power only a portion of atoms in the TC phase synchronize to the driving field, so that the global ensemble features a population of entrained atoms as well as an independently acting population undergoing phase diffusion. When the SIG field is turned off, the decay time, $\tau_{off}$, experimentally shows a slight increasing relationship with power (Fig. \ref{Fig:PowerSweep}d). In contrast, the classical model predicts $\tau_{off}$ is independent of $P$ during the off phase. The observed power dependence suggests that a phase-only model with constant diffusion coefficient is incomplete; suggesting some dynamics are not fully captured by our classical model. Adequately capturing the low power behavior of $\tau_{off}$ may therefore require a more involved model with additional amplitude or population dynamics.

In addition to the effect of SIG power, the entrainment is also dependent on beatnote detuning from the natural oscillation frequency. An example averaged trace for an 8.2 kHz beatnote applied to a TC natural oscillation frequency near 13.5 kHz shows a lack of entrainment following application of the SIG pulse (Fig. \ref{Fig:Detuning}a). In the frequency domain, the Lorentzian distributed natural oscillation frequencies can be seen along with the pulsed SIG field's Floquet sidebanded beatnote frequency, where the sideband spacing is equal to the pulse repetition frequency (Fig. \ref{Fig:Detuning}b-c). The apparent superposition of the Floquet sidebands along with the Lorentzian distributed time crystal oscillation peaks indicate that for the given pulsing sequence, the system exhibits characteristics of both a free running time crystal phase and heterodyned driving. Immediately following the SIG on phase, $\bar{A}_0^{off}$ can be seen to have a $\sim7$ kHz bandwidth for strong entrainment, followed by a rapid decay to a phase diffuse behavior (Fig. \ref{Fig:Detuning}d). The decay time of the synchronized oscillations, $\tau_{off}$, does not show a clear dependence on driving frequency (Fig. \ref{Fig:Detuning}e). At the strong driving conditions here, the system therefore shows a synchronization dependence that is dependent on the driving frequency, but the decay to a free running state retains no memory of the driving frequency.

\section{Discussion}

The time crystal oscillation phase for the three-level Rydberg atomic vapor system is frequently reported to experience phase noise and therefore, in the frequency domain, a distribution of frequencies around the natural oscillation frequency. External RF driving can modify the natural oscillation frequency distribution by shifting the frequency, causing injection locking, and/or leading to frequency comb formation \cite{manchaiah2026frequency}. Here, we use pulsed RF heterodyned driving to synchronize the ensemble oscillation phase and study the rise and fall dynamics. We observe that the oscillation decay time, $\gamma_{TC}$, is much longer than the standard decay time of an atom in a Rydberg state, $\gamma_{T}$. We additionally report on the effects of varying RF driving strength and detuning frequency with respect to the natural oscillation frequency. 

In order to model our experimental data, we developed a semiclassical model with driving based on the nonlinear Adler equation along with a phenomologically included phase diffusion term. The Adler equation is linked to the optical Bloch equations with a driving strength $K\propto\Omega_{rs} (n_r-n_s)$ where $n_r - n_s$ is the population inversion of the two Rydberg states \cite{arumugam2026injection}. A similar semiclassical approach is also presented in Ref. \cite{ cabot2024nonequilibrium} for a quantum van der Pol oscillator. We find that our model fits the time dynamics of the data in response to changing the heterodyned beat note frequency detuning and strength. Moreover, we note that we did not observe any subharmonic responses indicative of a discrete time crystal state under the periodic Floquet driving as was observed in Ref. \cite{liu2024higher}. Higher order harmonics were seen at multiples of the fundamental oscillation frequency. Therefore, the ensemble averaged coherence of the Rydberg CTC state exhibits a classical phase drift-diffusion behavior. However, our model does not provide a direct link between the quantum parameters that set the oscillation frequency, the driving strength, or the phase diffusion rates. Understanding this quantum to classical connection could therefore help in describing the complex nonequilibrium driven-dissipative nature of time crystal states.

One interesting question on Rydberg vapor time crystal states is what sets the decay time of the synchronized oscillatory state and why is the decay rate of the time crystal state, $\gamma_{TC}$, much less than the total decay rate of a standard EIT process, $\gamma_T$? One plausible model comes from an interesting study by Ding et al. \cite{ding2024ergodicity}. In this study, the authors numerically simulated a reduced-form master equation model of a spatially distributed population of two-level atoms with a mean-field coupling. The authors found that atoms in an oscillatory phase are spatially clustered and hypothesize that many-body interactions play a role in formation of the driven-dissipative oscillatory state. We find that our results are consistent with this hypothesis. In this model, a cluster of Rydberg atoms interacts via van der Waal's forces and experiences a single synchronized phase within the cluster. All the while, individual atoms in the cluster decay at a rate similar to the standard EIT configuration, $\gamma_T$. Nearby atoms pumped from the ground state to a Rydberg state undergoing time crystalline oscillations, which would normally have a free phase, instead synchronize their phase to the Rydberg cluster phase set by the many-body interactions. This many-body phase synchronization is therefore capable of extending the phase diffusion decay rate past the timescale expected from the normal Rydberg state decay times.

\section{Conclusion}

The hot-vapor Rydberg atom time crystalline order is a unique state of matter which exhibits long-time persistent oscillations. Many studies to date have reported phase diffusion and frequency instability in this system. The work here reports on rf heterodyne driving conditions which can entrain the phase and lead to long-time phase synchronization. We additionally show that a classical Fokker-Planck model with Adler-based driving describes the dynamical response. In total, this work contributes to time crystal applications requiring frequency and phase stability.

\section*{Acknowledgments}
The authors would like to thank Chang Li, Sean Muleady, and Alexey Gorshkov for helpful discussions. The views and opinions of authors expressed herein do not necessarily state or reflect those of the  United States Government or any agency thereof. A contribution of the U.S. government, this work is not subject to copyright in the United States.

\subsection*{Conflict of Interest}
\vspace{-3mm}
The authors have no conflicts to disclose.
\vspace{3mm}
\subsection*{Data Availability Statement}
\vspace{-3mm}
All data presented in this work is available to the reader at:

\bibliography{cites}
\end{document}